\documentclass[fleqn,5p]{elsarticle}

\journal{Applied Energy}

\usepackage{multicol}
\usepackage{amsmath}
\usepackage{amssymb}
\usepackage{mathtools}
\usepackage{cuted}
\usepackage{graphicx}
\usepackage{flafter}
\usepackage{placeins}
\usepackage{IEEEtrantools}
\usepackage{textcomp}
\usepackage{hyperref}
\usepackage{color}
\hypersetup{urlcolor=cyan}

\usepackage{flushend} 

\usepackage{etoolbox}
\makeatletter
\patchcmd{\ps@pprintTitle}
  {submitted to}
  {accepted for publication in}
  {}{}
\makeatother

\begin{document}

\setlength{\tabcolsep}{6pt}
\setlength{\mathindent}{0pt}

\begin{frontmatter}

\title{Ancillary services in Great Britain during the COVID-19 lockdown:\\a glimpse of the carbon-free future}

\author[imperial]{Luis Badesa}
\ead{luis.badesa@imperial.ac.uk}
\author[imperial]{Goran Strbac}
\ead{g.strbac@imperial.ac.uk}
\author[NationalGrid]{Matt Magill}
\ead{Matthew.Magill@nationalgrideso.com}
\author[NationalGrid]{Biljana Stojkovska}
\ead{Dr.Biljana.Stojkovska@nationalgrideso.com}
\address[imperial]{Imperial College London, Department of Electrical and Electronic Engineering, London SW7 2AZ, UK}
\address[NationalGrid]{National Grid Energy System Operator, Faraday House, Warwick Technology Park,
Warwick CV34 6DA, UK}

\begin{abstract}
The COVID-19 pandemic led to partial or total lockdowns in several countries during the first half of 2020, which in turn caused a depressed electricity demand. In Great Britain (GB), this low demand combined with large renewable output at times, created conditions that were not expected until renewable capacity increases to meet emissions targets in coming years. The GB system experienced periods of very high instantaneous penetration of non-synchronous renewables, compromising system stability due to the lack of inertia in the grid. In this paper, a detailed analysis of the consequences of the lockdown on the GB electricity system is provided, focusing on the ancillary services procured to guarantee stability. Ancillary-services costs increased by \pounds200m in the months of May to July 2020 compared to the same period in 2019 (a threefold increase), highlighting the importance of ancillary services in low-carbon systems. Furthermore, a frequency-secured scheduling model is used in the present paper to showcase the future trends that GB is expected to experience, as penetration of renewables increases on the road to net-zero emissions by 2050. Several sensitivities are considered, demonstrating that the share of total operating costs represented by ancillary services could reach 35\%. 
\end{abstract}


\end{frontmatter}

\vspace*{-6mm}
\vspace*{0.5mm}
\section*{\vspace*{0.5mm}Nomenclature}

\begin{tabular}{p{12mm} l}
\hspace{-3mm} $\eta_s^\textrm{c}$ & Charge efficiency for storage unit $s$.\\
\hspace{-3mm} $\eta_s^\textrm{d}$ & Discharge efficiency for storage unit $s$.\\
\hspace{-3mm} $\pi(n)$ & Probability of reaching node $n$ in the \\ & stochastic scheduling.\\
\hspace{-3mm} $\Delta f_{\mathrm{max}}$ & Maximum admissible frequency deviation \\ & at the nadir (Hz).\\
\hspace{-3mm} BESS & Battery Storage Energy Systems. \\
\hspace{-3mm} $\textrm{BECSS}$ & Bio-Energy with Carbon Capture and \\ & Storage.\\
\hspace{-3mm} $\textrm{CCS}$ & Carbon Capture and Storage.\\
\hspace{-3mm} $\textrm{CfD}$ & Contract for Difference.\\
\hspace{-3mm} $C_g$ & Operating cost of generator $g$ (\pounds).\\
\hspace{-3mm} $\mathrm{c}_{g}^{\mathrm{st}}$ & Startup cost of generator $g$ (\pounds).\\
\hspace{-3mm} $\mathrm{c}_{g}^{\mathrm{nl}}$ & No-load cost of generator $g$ (\pounds/h).\\
\hspace{-3mm} $\mathrm{c}_{g}^{\mathrm{m}}$ & Marginal cost of generator $g$ (\pounds/MWh).\\
\hspace{-3mm} $\textrm{D}$ & Total system demand (MWh).\\
\hspace{-3mm} $\textrm{D}_\textrm{shed}$ & Demand shed (MWh).\\
\hspace{-3mm} $E_s$ & State of charge of storage unit $s$ ($\textrm{MWh}$).\\
\end{tabular}

\begin{tabular}{p{12mm} l}
\vspace*{8.5mm}\\
\hspace{-3mm} $\textrm{E}_s^\textrm{max}$ & Maximum charge level for storage unit $s$ \\ & ($\textrm{MWh}$).\\
\hspace{-3mm} $\textrm{E}_s^\textrm{min}$ & Minimum charge level for storage unit $s$ \\ & ($\textrm{MWh}$).\\
\hspace{-3mm} $\operatorname{EFR}$ & Enhanced Frequency Response ($\textrm{MW}$).\\
\hspace{-3mm} $\operatorname{EFR}_s$ & EFR delivered by storage unit $s$ ($\textrm{MW}$).\\
\hspace{-3mm} $\operatorname{EFR}_s^\textrm{max}$ & EFR capacity of storage unit $s$ ($\textrm{MW}$).\\
\hspace{-3mm} $f_0$ & Nominal frequency of the power grid (Hz).\\
\hspace{-3mm} $g$ & Index for thermal generators.\\
\hspace{-3mm} $\operatorname{GB}$ & Great Britain.\\
\hspace{-3mm} $H$ & Total system inertia ($\textrm{MVA}\cdot \textrm{s}$).\\
\hspace{-3mm} $\textrm{H}_g$ & Inertia constant of generator $g$ ($\textrm{s}$). \\   
\hspace{-3mm} $\textrm{H}_\textrm{Loss}$ & Inertia constant of outaged unit ($\textrm{s}$). \\   
\hspace{-3mm} $n$ & Index for nodes in the scenario tree of \\ & the stochastic scheduling. \\
\hspace{-3mm} $\textrm{OCGT}$ & Open-Cycle Gas Turbine.\\
\hspace{-3mm} $\textrm{OFDM}$ & Optional Downward Flexibility
\\ & Management.\\
\hspace{-3mm} $\operatorname{PFR}$ & Primary Frequency Response ($\textrm{MW}$).\\
\hspace{-3mm} $\textrm{PFR}_g$ & PFR delivered by generator $g$ ($\textrm{MW}$).\\
\hspace{-3mm} $\textrm{PFR}_g^\textrm{max}$ & PFR capacity of generator $g$ ($\textrm{MW}$).\\
\hspace{-3mm} $\operatorname{PHES}$ & Pumped Hydroelectric Energy Storage.\\
\hspace{-3mm} $P_s^\textrm{c}$ & Charge rate of storage unit $s$ ($\textrm{MW}$).\\
\end{tabular}

\begin{tabular}{p{15mm} l}
\hspace{-3mm} $\textrm{P}_s^\textrm{c, max}$ & Maximum charge rate for storage unit \\ & $s$ ($\textrm{MW}$).\\
\hspace{-3mm} $P_s^\textrm{d}$ & Discharge rate of storage unit $s$ ($\textrm{MW}$).\\
\hspace{-3mm} $\textrm{P}_s^\textrm{d, max}$ & Maximum discharge rate for storage \\ & unit $s$ ($\textrm{MW}$).\\
\hspace{-3mm} $P_g$ & Power output of generator $g$ ($\textrm{MW}$).\\
\hspace{-3mm} $\textrm{P}_g^\textrm{max}$ & Rated power of generator $g$ ($\textrm{MW}$).\\
\hspace{-3mm} $\textrm{P}_g^\textrm{msg}$ & Minimum stable generation of generator \\ & $g$ ($\textrm{MW}$).\\
\hspace{-3mm} $\textrm{P}_g^\textrm{rr}$ & Maximum ramp rate of generator $g$ \\ & ($\textrm{MW}/\textrm{h}$).\\
\hspace{-3mm} $P_\textrm{Loss}$ & Largest power infeed loss ($\textrm{MW}$).\\
\hspace{-3mm} $\textrm{P}_\textrm{Loss}^\textrm{max}$ & Rated power of the largest unit ($\textrm{MW}$).\\
\hspace{-3mm} $\textrm{P}_\textrm{RES}$ & RES power output ($\textrm{MW}$).\\
\hspace{-3mm} $P_\textrm{RES}^\textrm{curt}$ & RES power curtailed ($\textrm{MW}$).\\
\hspace{-3mm} $\textrm{RES}$ & Renewable Energy Sources.\\
\hspace{-3mm} $\operatorname{RoCoF}$ & Rate-of-Change-of-Frequency.\\
\hspace{-3mm} $\textrm{RoCoF}_\textrm{max}$ & Maximum admissible RoCoF (Hz/s).\\
\hspace{-3mm} $\textrm{T}_g^\textrm{mdt}$ & Minimum down time for generator $g$ (h).\\
\hspace{-3mm} $\textrm{T}_g^\textrm{mut}$ & Minimum up time for generator $g$ (h).\\
\hspace{-3mm} $\textrm{T}_g^\textrm{st}$ & Startup time for generator $g$ (h).\\
\hspace{-3mm} UC & Unit Commitment. \\
\hspace{-3mm} VoLL & Value of Lost Load (\pounds/MWh). \\
\hspace{-3mm} $y_s$ & Charge/discharge state variable for \\ & storage unit $s$.\\
\hspace{-3mm} $y_g$ & Commitment state of generator $g$.\\
\hspace{-3mm} $y_g^\textrm{sd}$ & Shutdown variable for generator $g$.\\
\hspace{-3mm} $y_g^\textrm{sg}$ & `Start generating' state of generator $g$.\\
\hspace{-3mm} $y_g^\textrm{st}$ & Startup variable for generator $g$.\\
\end{tabular}

\section{Introduction}

The outbreak of the disease COVID-19 in the final months of 2019 had a significant impact on electricity systems worldwide. Partial or total lockdowns were enforced by governments of several states, starting in China, followed by Italy and several European countries, as well as the United States. The measures taken varied from country to country, with several governments forbidding citizens from leaving their homes except for essential activities, and most of the industrial activity being put on halt. In Great Britain (GB), these measures caused the electricity demand to drop by up to 28\% compared to levels in the same period in previous years \cite{DemandCovidNG}.

Several studies have discussed the effects of lockdowns due to COVID-19 on electricity systems. Most of the focus has been put on the reduction in demand \cite{LopezProlCovid}, with the subsequent drop in energy prices and carbon emissions \cite{IEA_Covid}, since demand in many countries was mostly covered by emissions-free renewables, which also benefit from low short-run marginal costs. Reference \cite{ArticleCovid1} has documented the correlation between
reduction in electricity consumption in the United States and the rise in number of COVID-19 cases, degree of social distancing, and level of commercial activity.
The work in \cite{EU_ETS_Covid} analyses the sharp drop of almost 40\% in prices in the European carbon market in March 2020, driven by the surplus of emissions allowances as thermal generators were largely pushed out of electricity markets due to low demand. The authors in \cite{KeithBellCovid} analysed the demand reduction in Great Britain during the first week of the lockdown, highlighting some of the challenges that the system operator was expected to experience.
The present paper focuses on another important aspect, that will only increase in importance as energy systems shift towards net-zero emissions: ancillary services. 

A main challenge caused by low demand is related to grid stability: maintaining the network magnitudes, such as voltage and frequency, within safe levels is far from being trivial given these exceptional conditions. The authors in \cite{KangCovid} discuss the preventive reactive-power management undertaken by the Indian system operator to avoid potentially dangerous voltage rises in the network. Such voltage rises could have been caused by an excess of reactive power, as there was a decrease in load while distributed renewable generation output did not decrease. In Great Britain, an island with no synchronous interconnection to other countries, low levels of inertia driven by high penetrations of non-synchronous renewable generation are a concern, which require large volumes of ancillary services to guarantee stability in the event of a contingency. A recent report \cite{StaffellCovid} focused on the British system, highlighting that a significant proportion of demand in GB has been covered at times from non-synchronous renewables such as wind and solar PV, and providing a summary of balancing costs. However, no publication to date has provided a comprehensive analysis of the challenges experienced in GB, or the specific actions taken by the system operator and the reasoning behind them.

Furthermore, the lessons learned in GB are of uttermost relevance for the future decarbonised electricity grids. Some works such as \cite{CCCreportImperial} have already quantified the future cost of ancillary services, demonstrating that these will represent a very significant share of total operating costs for the electricity grid. However, an updated analysis is necessary given two major recent changes: 1) the ambition to sharply reduce emissions has greatly increased recently: as an example, the British system operator National Grid envisions that net emissions from the power sector will be negative by 2033 in 3 out of 4 future scenarios (achieved by including certain capacity of bio-energy with carbon capture) \cite{NationalGridFES2020}.
And 2) previous modelling tools did not have the capabilities to account for fast frequency response, which has been demonstrated as key to tackle the low-inertia problem (with 1GW of fast response expected to be procured by mid-2021 in GB \cite{DynamicContainment2020}). 

Several works have proposed optimisation models to schedule operating reserves \cite{AlexEfficient}, frequency response from thermal units \cite{OPFChavez} and synchronous inertia \cite{FeiStochastic}. While frequency response is already co-optimised along with energy in some countries like the US, only the quasi-steady-state value is considered (i.e.~the same volume of frequency response is scheduled as MW’s would be lost after the largest generator outage). The quasi-steady-state value of frequency after a contingency was the only concern until recent years due to the high level of inertia in the grid, provided by synchronous generators as a by-product of energy. Nevertheless, decreasing inertia levels imply that a co-optimisation of inertia and response is necessary to compute the required volume of these services, as demonstrated by \cite{FeiStochastic}. Furthermore, only recently has grid-scale battery storage become a cost-effective option, and this technology is the main provider of ‘fast frequency response’ in grids such as GB. The recent development of modelling tools that allow to also co-optimise fast frequency response \cite{LuisMultiFR} make it possible to analyse ancillary services requirements under the latest expectations for future generation mix.

In this context, the present work analyses the challenge of procuring ancillary services in GB during the COVID-19 lockdown, and demonstrates how these challenges will also be present in the future system as it becomes decarbonised. The contributions of this paper are two-fold:
\begin{enumerate}
    \item To provide a comprehensive analysis of the stability challenge faced by the British electricity system during the COVID lockdown, as well as discussing the additional ancillary services procured by the system operator in this period and their associated costs.
    \item To quantify the future costs for stability actions in GB and analyse their main drivers. To do so, a frequency-secured scheduling model is used, which explicitly considers the stability boundary within the optimisation. Several sensitivities are considered, demonstrating the importance of fast frequency response and size of the largest possible contingency.
\end{enumerate}

The remainder of this paper is organised as follows: Section~\ref{Sec:CovidGB} reviews the need for and types of ancillary services procured by the GB system operator to guarantee stability during the lockdown period. Section~\ref{Sec:FutureTrends} illustrates why the lessons learned during this period can inform the challenges that will be faced by the future high-renewable grids, using several results from a frequency-secured scheduling model that co-optimises energy production and ancillary services. Finally, Section~\ref{Sec:Conclusion} provides the conclusion and discusses future lines of work needed to enhance our understanding of best practices in ancillary-services procurement.

\vspace{-1.2mm}

\section{Ancillary services in GB during COVID-19} \label{Sec:CovidGB}

In this section a discussion on the operational challenges experienced in GB during the lockdown period is provided, as well as the actions taken by the system operator to overcome them. Finally, the consequences of this unique period are explained, both in terms of economic and environmental impact.

\subsection{Challenges experienced} \label{Sec:Challenges}

The challenges faced by the British electricity system during the lockdown period were due to three factors: 
\begin{enumerate}
    \item Depressed electricity demand.
    \item Most renewables are not affected by market prices, due to their special market arrangements.
    \item Nuclear stations are also not significantly affected by market prices, due to their inflexibility.
\end{enumerate}

The first and second factors above caused periods of low net-demand (i.e.~demand minus renewable generation, which must be covered by dispatchable generation), which in turn caused low system inertia: inertia is provided by synchronous generators (mainly gas- and coal-fired power stations), which were displaced by non-synchronous renewables such as wind and solar PV. The challenge of operating a low-inertia grid, which is further explained in Section~\ref{Sec:ConditionsFrequency}, means that a higher volume of alternative ancillary services is needed to guarantee system stability.

The drop in demand caused a drop in energy prices, but most renewables in GB are incentivised to produce as much energy as possible regardless of the energy price. The reason for this is not only the low short-run marginal costs of these technologies (given that they do not incur fuel costs),
but also their special market arrangements in GB: many renewable generators benefit from special financial instruments such as feed-in tariffs, Renewable Obligation Certificates and Contracts for Difference (CfDs). Although these three instruments vary in their characteristics, they all share a common feature: they shield renewables from low market prices, allowing these generators to still profit from generating even during periods of negative energy prices. A CfD, which is the financial support currently being awarded by the UK government to renewable generators successful in CfD auctions, guarantees a fixed strike price per MWh of energy paid to the generator, regardless of the price set in the wholesale market. This strike price effectively incentivises the renewable producer to output as much energy as possible, making it profitable even in the event of negative energy prices (since generators under CfDs are currently allowed to bid negatively in the wholesale market).

Simultaneously, the third factor contributing to the operational challenge during lockdown was that large nuclear units were predicted to maintain a high load factor, which would have increased the need for ancillary services. The reason for this increased need for ancillary services is that these units, in particular station Sizewell~B, typically drive the largest power infeed loss in GB: since the system operator enforces the `\textit{N}-1' reliability requirement (meaning that the grid must withstand any single failure of a device), there must be enough ancillary services to cover for the outage of the largest plant online at any time.

Nuclear plants were likely to maintain high load factors because they are somewhat indifferent to low energy prices: given the inflexibility of these units in GB (long commitment times and high startup costs), it is typically preferable for these plants to stay online through periods of low prices rather than turn off. 

In summary, the periods of very low net-demand combined with high nuclear output increased the need for ancillary services in GB during the COVID-19 lockdown. As an illustrative example, the lowest ever observed national demand 
occurred during the lockdown: 13.4GW, overnight on Sunday June 28\textsuperscript{th} 2020 \cite{NGlowestDemand}. 
To compare with 2019, the lowest demand was of roughly 18GW that year. 
In the next section, a discussion on the actions taken by the system operator in GB to overcome this operational challenge is presented.

\subsection{Measures taken} \label{Sec:MeasuresTaken}

The British government enforced a lockdown some days after other European countries had: lockdown in the United Kingdom started on 23\textsuperscript{rd} March 2020, while in Spain it had started on 13\textsuperscript{th} March and in the Lombardy region of Italy restrictions had been in place since around 21\textsuperscript{st} February. This delay gave some valuable days to the GB system operator, National Grid, to conduct studies on how low demand could drop during coming months, since there was some concern were the lockdown to extend into the summer (which is the period of lowest demand in GB). By early April 2020, roughly a week after the lockdown was enforced, National Grid had developed a projection for the expected levels of demand in coming months: demand was expected to drop by up to 20\% compared to the same periods in previous years \cite{SummerOutlook2020}.

These studies conducted by National Grid foresaw periods when it would be impossible to guarantee system stability. The system operator estimates that the capacity of synchronous units needed online to guarantee stability is in the order of 8 to 9GW, while the lowest possible demand foreseen during the summer was just above 12GW.
Given that the installed renewable generation and about 5GW of nuclear power stations on the system were expected to generate as usual, as they are somewhat indifferent to demand drops (explained in Section~\ref{Sec:Challenges}), National Grid realised that there was a key necessity: the system was going to require more downward flexibility \cite{NGpodcastCovid19}.

To obtain this downward flexibility, National Grid created a new product, called Optional Downward Flexibility Management (OFDM). The goal was to either increase the demand of the transmission system, or decrease the amount of non-flexible providers outputting onto the system. This OFDM product provided a commercial route for providers with whom National Grid did not have a commercial flexibility arrangement with: over 4.5GW of assets signed up for this service, predominantly embedded wind and solar, but also a fair share of demand turn-up assets.

The OFDM service was used 4 times, with a total cost of the actions of approximately \pounds7m \cite{NGpodcastCovid19}.
These instructions were used both during daytime (such as from 6am to 4.30pm due to high solar output) and nighttime periods (as demand is lowest during night, which was combined with high wind in these periods).
It is notably illustrative that three of these four OFDM instructions happened on the second bank-holiday weekend in the month of May (22\textsuperscript{nd} to 25\textsuperscript{th} of May 2020), since demand is traditionally lowest in GB during bank holidays and weekends.

Another measure taken by the system operator was to sign a bilateral contract with the nuclear power station Sizewell~B. Through this contract, National Grid can instruct Sizewell~B to halve its output: given that this is the largest generator in GB (with 1.3GW rated power), it defines the needed volume of ancillary services through the `\textit{N}-1' reliability requirement; therefore, part-loading this unit effectively reduces the need for ancillary services. The cost of this bilateral contract has been reported to be of up to \pounds73m \cite{TimesSizewellB}.

Furthermore, National Grid also takes actions to guarantee system stability by committing out-of-merit synchronous generators in the balancing market, called the Balancing Mechanism in GB. As an example, 2.3GW of gas plants were brought online through the Balancing Mechanism on the Easter Monday bank holiday (13\textsuperscript{th} April 2020) with the purpose of increasing inertia and procuring additional frequency capability \cite{BM_actions_April2020}.

Finally, the last measure taken was to create a fast-track application process to the Accelerated Loss of Mains Change Programme \cite{LossOfMainsProgramme}, targeted to embedded resources. Through this fast-track programme, generators receive payments from National Grid to change their protection settings, so that they stay connected to the grid under high Rate-of-Change-of-Frequency conditions. The programme will be completed in coming years, but the fast track created during the lockdown had the goal of decreasing the need for one key ancillary service that was scarce during this period of low net-demand: inertia.

In summary, the GB system operator took four different measures to avoid blackouts during the challenging lockdown period: 
\begin{enumerate}
    \item To create the OFDM product in order to procure downward flexibility.
    \item To sign a bilateral contract with the large power station Sizewell~B, allowing to halve its power output.
    \item To take stability-related actions through the Balancing Mechanism, as usual practice before the lockdown.
    \item To create a fast-track route for the Accelerated Loss of Mains Change Programme, which will reduce the need for inertia.
\end{enumerate}

\subsection{Economic and environmental consequences}

The actions to procure additional ancillary services explained in Section~\ref{Sec:MeasuresTaken} implied additional costs. Ancillary-services costs were 3 times higher in the period May-July 2020 compared to the same months in 2019: £302m in 2020 compared to £101m in 2019 \cite{AScosts2020}.

On the positive side, the high penetration of renewable generation during this period has produced records of minimal carbon intensity: May 2020 was the lowest carbon-intensity month for electricity so far in Great Britain, with May 24\textsuperscript{th} showing the lowest ever carbon intensity of the system (46gCO\textsubscript{2}/kWh produced on the transmission network) \cite{NGpodcastCovid19}. One of the reasons for this drop in carbon emissions during lockdown was the halt in coal usage: the UK did not use any coal for power generation for just under 68 consecutive days, from April to June 2020. This set a record for the longest coal-free period since the industrial revolution, exceeding by far the previous record of 18 days achieved in 2019 \cite{NoCoalRecord2020}.

\section{Future trends} \label{Sec:FutureTrends}

Even though electric demand is expected to increase in GB in coming decades due to the electrification of heat and transport, net-demand is expected to further decrease due to an even higher rate of increase of renewable penetration to meet the emission targets \cite{NationalGridFES2020}. In addition, an increase in the largest power infeed loss with the commissioning of large nuclear plants such as Hinkley Point C in coming years \cite{HinkleyPointC}, will increase the need for ancillary services to maintain system stability.

\begin{figure} [t]
    \centering
    \includegraphics[width=3.3in]{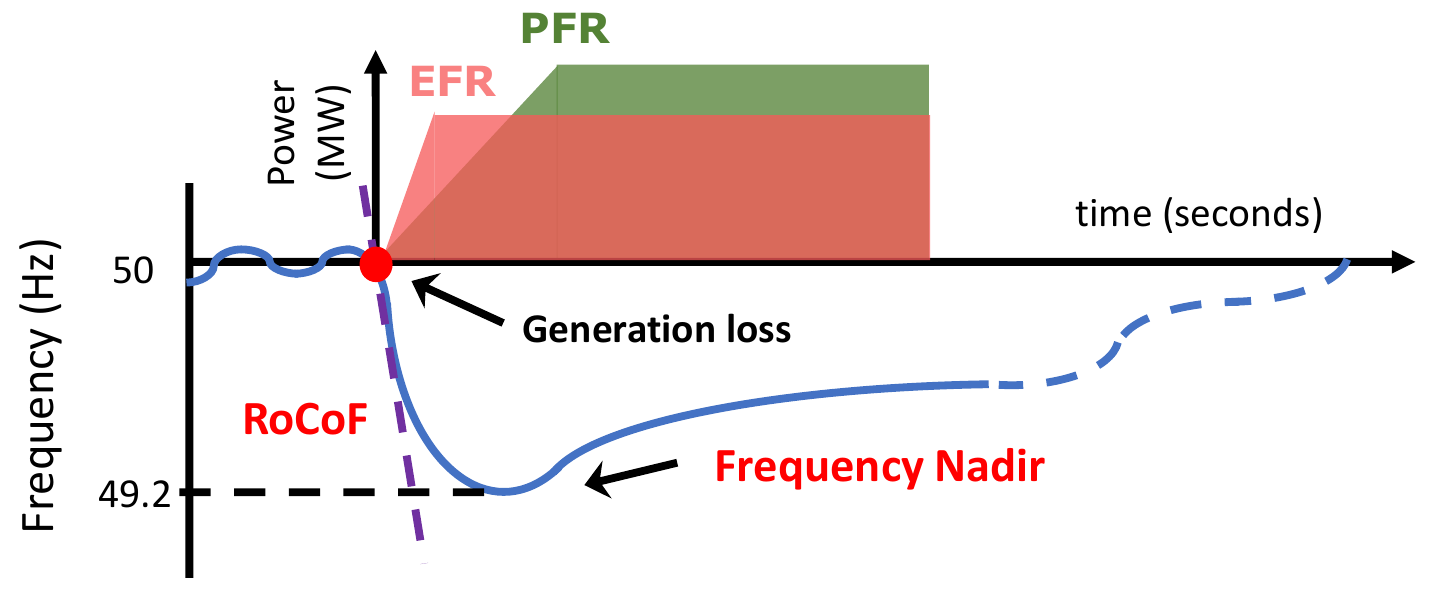}
    \caption{Post-contingency frequency evolution and the limits that must be respected in GB, along with the power contribution from the two frequency-response services:  Enhanced Frequency Response (EFR) and Primary Frequency Response (PFR).}
	\label{fig:FrequencyGraph}
\end{figure}

These two effects, the low inertia caused by low net-demand
and the high value of the largest power infeed, will exacerbate in coming decades the operational challenge already experienced during the lockdown of 2020. To illustrate this point, a frequency-secured scheduling tool is used in this section considering National Grid's latest view on the generation mix of the future, demonstrating that the cost of ancillary-services provision will increase as GB decarbonises its grid to meet the target of net-zero carbon emissions by 2050.

Note that the present analysis focuses on ancillary services to guarantee frequency stability, which is the main concern in low-inertia power systems. With increasing loading level in the grid, voltage stability could also be compromised. Other concerns could also arise, such as a higher difficulty in respecting rotor-angle stability, and even new types of potential instability driven by the fast dynamics of power-electronics-based generation \cite{IEEEstabilityClassification}. These types of stability are not considered in this paper.

\subsection{Conditions for frequency stability} \label{Sec:ConditionsFrequency}

National Grid, as the system operator in Great Britain, must take actions to guarantee that the frequency of the grid is kept within safe boundaries after the loss of any generator (an operational practice commonly known as the `\textit{N}-1 reliability requirement'). These safe boundaries are defined as:
\begin{enumerate}
    \item The Rate-of-Change-of-Frequency (RoCoF) must not exceed 0.125Hz/s, in order to avoid triggering  RoCoF-sensitive relays that could further exacerbate the generation loss by disconnecting distributed generation. This RoCoF limit is currently being relaxed to 0.5Hz/s and 1Hz/s in certain locations through the Accelerated Loss of Mains Change Programme, as the cost of RoCoF-related ancillary services is expected to significantly increase in coming years if this limit is not relaxed \cite{NationalGridRelaxRocof}.
    
    \item The lowest frequency value, the frequency nadir, must not drop below 49.2Hz. This boundary avoids the activation of Low Frequency Demand Disconnection, which would entail significant economic penalties for the parties responsible, as occurred after the large power outage of August 9\textsuperscript{th} 2019 \cite{OfgemOutage2019}.
\end{enumerate}

In order to guarantee that the two above frequency boundaries are respected, the energy market must be constrained so that the generation dispatch contains sufficient ancillary services. These ancillary services for frequency support in GB are: 1) inertia; 2) Enhanced Frequency Response (EFR); and 3) Primary Frequency Response (PFR) \cite{FutureBalancingServices}.
These ancillary services help contain the frequency drop that would be caused by a generation outage, as described in Figure~\ref{fig:FrequencyGraph}.

Mathematically, the conditions to guarantee frequency stability take the following shape. The RoCoF is guaranteed to be within safe limit if the following inequality holds
\cite{NGrocofDefinition}: 
\begin{equation} \label{eq:RoCoF}
    \frac{H}{f_0} \geq \frac{P\textsubscript{Loss}}{2 \cdot \textrm{RoCoF}\textsubscript{max}}
\end{equation}
where $H$ is the system inertia provided by all synchronous generators online (with units of $\textrm{GVA}\cdot\textrm{s}$), $f_0$ is the nominal frequency of the grid (50Hz in European grids), $P\textsubscript{Loss}$ is the largest power infeed loss (units of $\textrm{GW}$) and $\textrm{RoCoF}\textsubscript{max}$ is the maximum admissible RoCoF (units of Hz/s).

The frequency nadir is guaranteed to be above the pre-defined limit if the following inequality is respected \cite{LuisMultiFR}:
\begin{equation} \label{eq:nadir}
    \left(\frac{H}{f_0} - \frac{\textrm{EFR}\cdot1\textrm{s}}{4\cdot \Delta f\textsubscript{max}}\right)\cdot \frac{\textrm{PFR}}{10\textrm{s}} \geq \frac{(P\textsubscript{Loss} - \textrm{EFR})^2}{4\cdot \Delta f\textsubscript{max}}
\end{equation}
where $\Delta f\textsubscript{max}$ is the maximum admissible frequency deviation (in Hz), EFR is the aggregate from all devices providing this service (units of MW, EFR must be fully delivered by 1s after the contingency), and PFR is the aggregate from all devices providing this service (units of MW, PFR must be fully delivered by 10s after the contingency).

Note that the EFR service is currently being migrated into a service with similar characteristics named Dynamic Containment \cite{FutureResponseServices2020}.

\begin{table*}[!t]
{%
\centering
\setlength\tabcolsep{10pt} 
\begin{tabular}[\textwidth]{l|cccccc}
\hline
\hline
\multicolumn{1}{c|}{}    & Hinkley Point C   & Nuclear & Gas CCS  & OCGT   & Biomass & BECSS  \\
\hline
Number of Units            & 1 & 2       & 45    & 10     & 9   & 7 \\
Rated Power (MW)           & 1800   & 1350  & 500   & 100    & 500  & 500 \\
Min Stable Generation (MW) & 1350   & 1000  & 250 & 50     & 450  & 450 \\
Ramp Rate (MW/h) & 100   & 100  & 250 & 50     & 50  & 50 \\
No-Load Cost (\pounds/h)         & 400   & 400  & 4500  & 3000  & 5000 & 3500  \\
Marginal Cost (\pounds/MWh)      & 8  & 8  & 46 & 200 & 25 & 15  \\
Startup Cost (\pounds'000)           & N/A & N/A & 10 & 0      & 20 & 20 \\
Startup Time (h)           & N/A & N/A     & 4     & 0      & 4   & 4  \\
Min Up Time (h)            & N/A & N/A     & 4     & 0      & 4    & 4 \\
Min Down Time (h)          & N/A & N/A     & 1      & 0      & 1   & 1    \\
Inertia Constant (s)       & 5       & 5     & 5      & 5   & 5  & 5  \\
Max PFR deliverable (MW)   & 0       & 0       & 50   & 20     & 0  & 0 \\
\hline
\hline
\end{tabular}
}
\caption{\label{tab:GenerationMix} Characteristic of thermal plants} 
\end{table*}

\subsection{Frequency-secured scheduling} \label{Sec:SchedulingModel}

In order to understand the need for ancillary services in the future carbon-free GB grid, several relevant case studies are run here using a frequency-secured stochastic Unit Commitment model with two important characteristics: 1) Considers uncertainty from renewable generation, by using a scenario tree of possible RES realisations; and 2) Guarantees frequency stability, by enforcing frequency constraints. 

The objective function of this frequency-secured scheduling minimises the expected value of system operating costs (fuel costs, start-up costs) to meet demand:
\begin{equation} 
\text{min}\quad \sum_{\forall n}\pi(n) \left( \sum_{\forall g}C_g(n) + \textrm{VoLL}\cdot \textrm{D}_\textrm{shed}(n) \right)
\end{equation}
Where the operating cost of thermal units is given by:
\begin{equation} 
C_g(n)=\textrm{c}_g^\textrm{st}\cdot y_g^\textrm{sg}(n) + \textrm{c}_g^\textrm{nl}\cdot y_g(n) + \textrm{c}_g^\textrm{m}\cdot P_g(n)
\end{equation}

The objective function is subject to engineering constraints in the generation mix, such as start-up times and ramp rates for thermal plants. The full description of these constraints can be found in \ref{app:UC}. The binary variables in the UC are relaxed to continuous in order to decrease computation time. The objective function is also subject to the frequency-stability constraints~(\ref{eq:RoCoF}) and~(\ref{eq:nadir}), to guarantee that the scheduling solution would allow to keep the RoCoF and frequency nadir within required limits in the event of any credible contingency.

The frequency-stability constraints in Section~\ref{Sec:ConditionsFrequency} map the sub-second dynamics of post-contingency frequency evolution (shown in Figure~\ref{fig:FrequencyGraph}), into any timescale desired for the optimisation, such as the hourly timescale used in this scheduling model. These constraints make sure that sufficient ancillary services will be available to contain any credible frequency drop: inertia, EFR and PFR are decision variables in the scheduling optimisation.

Inertia refers to the energy stored in the rotating masses on synchronous generators (e.g.~nuclear, gas, biomass). These masses act as an energy buffer that stores kinetic energy as they rotate: this energy is spontaneously released in the event of a generation outage, decreasing the rotating speed of these masses and therefore decreasing the electrical frequency of the grid. A higher number of synchronous generators committed increases system inertia.

Frequency Response is a power injection from different assets such as generators and batteries. Generators have slower dynamics in frequency-response provision, making them candidates for the PFR service (which has a requirement of being fully delivered by 10s after a contingency). Batteries and other devices controlled through power electronics converters have much faster dynamics, allowing them to provide EFR (which must be fully delivered by 1s after a contingency).

All these ancillary services can be thought of as a form of `insurance' to prevent blackouts: for example, part-loaded generators providing frequency response operate at a lower-than-optimal power output, therefore increasing their per MWh operating cost. The headroom in part-loaded generators provides this insurance in the form of frequency response, that will act if a generation outage occurs.
Another example of this insurance consists in bringing out-of-merit thermal units to increase the system inertia: these thermal units brought online displace non-synchronous renewables, therefore increasing the fuel costs in the system, as demonstrated in detail in \cite{LuisPricing}.

\subsection{Ancillary services costs by 2030} \label{Sec:Results}

\begin{figure} [!t]
\hspace*{-1.5mm}
    \includegraphics[width=3.3in]{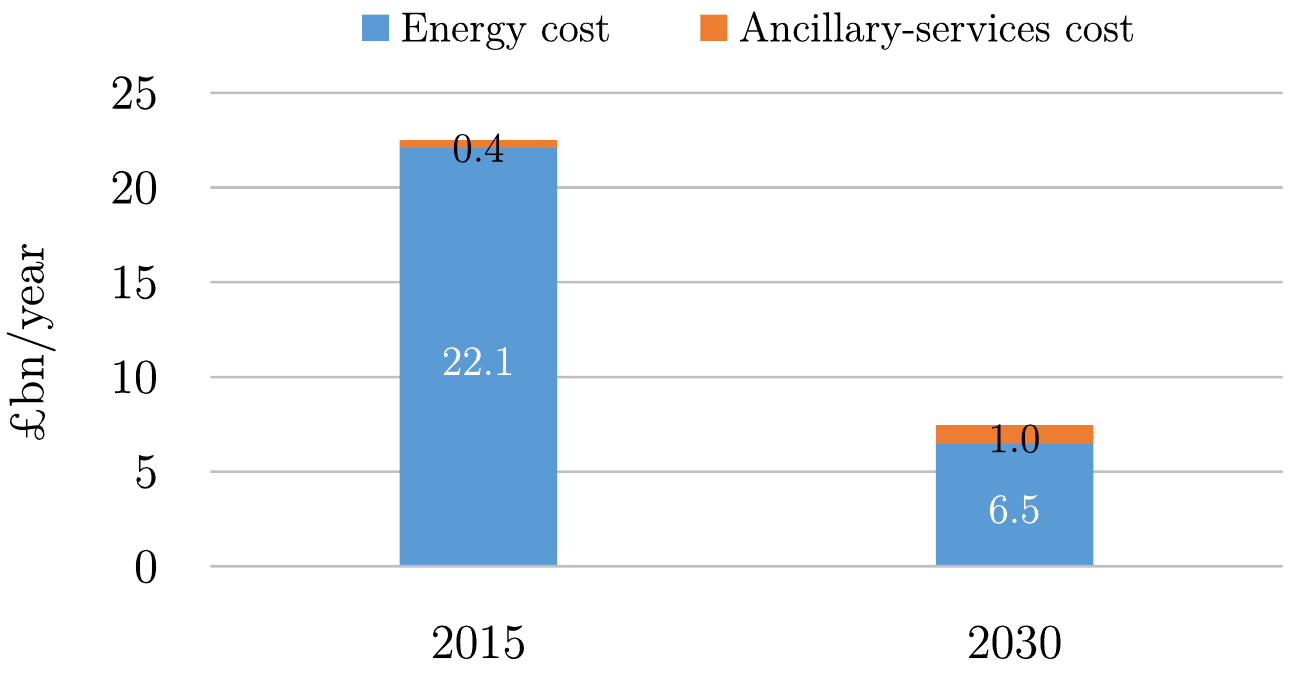}
    \caption{Energy and ancillary services costs in GB in 2015 \cite{CCCreportImperial} and projected costs in 2030 for National Grid's `Leading the way' scenario.}
	\label{fig:ResultsSUC}
\end{figure}

The frequency-secured scheduling introduced in Section~\ref{Sec:SchedulingModel} is used here to understand the future trends on ancillary services provision. The generation mix considered is described in Table~\ref{tab:GenerationMix}, and corresponds to the `Leading the way' scenario within National Grid's 2020 Future Energy Scenarios \cite{NationalGridFES2020}, described as the `fastest credible path to decarbonisation'. This scenario achieves a carbon-neutral electricity system by 2030, and is the result of a capacity-expansion optimisation which considers a set of different technologies as generation candidates, subject to regulatory constraints such as emissions limits. 

Electric demand has an annual peak of 60GW and a minimum value of 20GW \cite{NationalGridFES2020}, accounting for daily and seasonal trends. 
The RES fleet is composed of 30GW of solar PV and 68GW of wind. In addition, the system contains 14 GW of storage, including the 2.8GW existing Pumped Hydroelectric Energy Storage (PHES) plus a further 2GW assumed to be built by 2030, and 9.2GW of Battery Energy Storage Systems (BESS). PHES has a round-trip efficiency of 75\% and 5h duration, while BESS has a round-trip efficiency of 90\% and 2h duration. In the base case, 1GW of EFR is procured, as per current plans by National Grid from mid-2021 onwards \cite{DynamicContainment2020}.

The results for a scheduling simulation spanning one full year of operation are presented in Figure~\ref{fig:ResultsSUC}. The cost of ancillary services is computed by running two simulations, one with the frequency-stability constraints~(\ref{eq:RoCoF}) and~(\ref{eq:nadir}) enforced, and another one without these constraints (therefore corresponding to an energy-only scheduling). The difference between the operating cost in the first simulation and the operating cost in the second simulation gives the cost of ancillary-services provision.

Figure~\ref{fig:ResultsSUC} illustrates that the share of total system operating costs corresponding to ancillary services would rise to 15\% by 2030 under the `Leading the way' scenario. For comparison, Figure~\ref{fig:ResultsSUC} also displays the estimated energy and ancillary services costs in 2015, when only 2\% of costs corresponded to ancillary services \cite{CCCreportImperial}.

\begin{figure} [!t]
\hspace*{-2mm}
    \centering
    \includegraphics[width=3.3in]{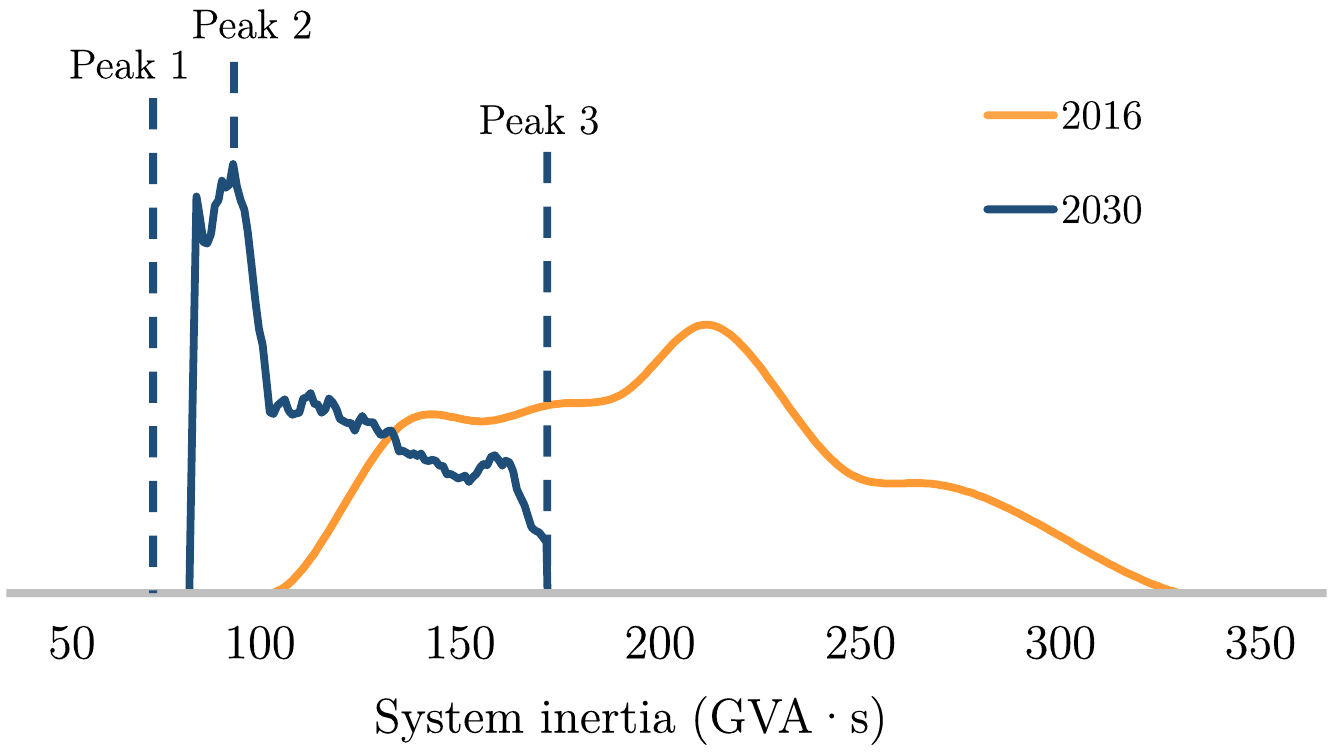}
    \caption{Annual distribution of system inertia for 2016 \cite{NGrocofDefinition} and projected distribution for 2030 under National Grid's `Leading the way' scenario. The dashed lines represent peaks of the 2030 distribution that correspond to a single value of system inertia and fall outside the scale of the graph.}
	\label{fig:inertia_distribution}
\end{figure}

\begin{figure} [!t]
    \centering
    \includegraphics[width=3.3in]{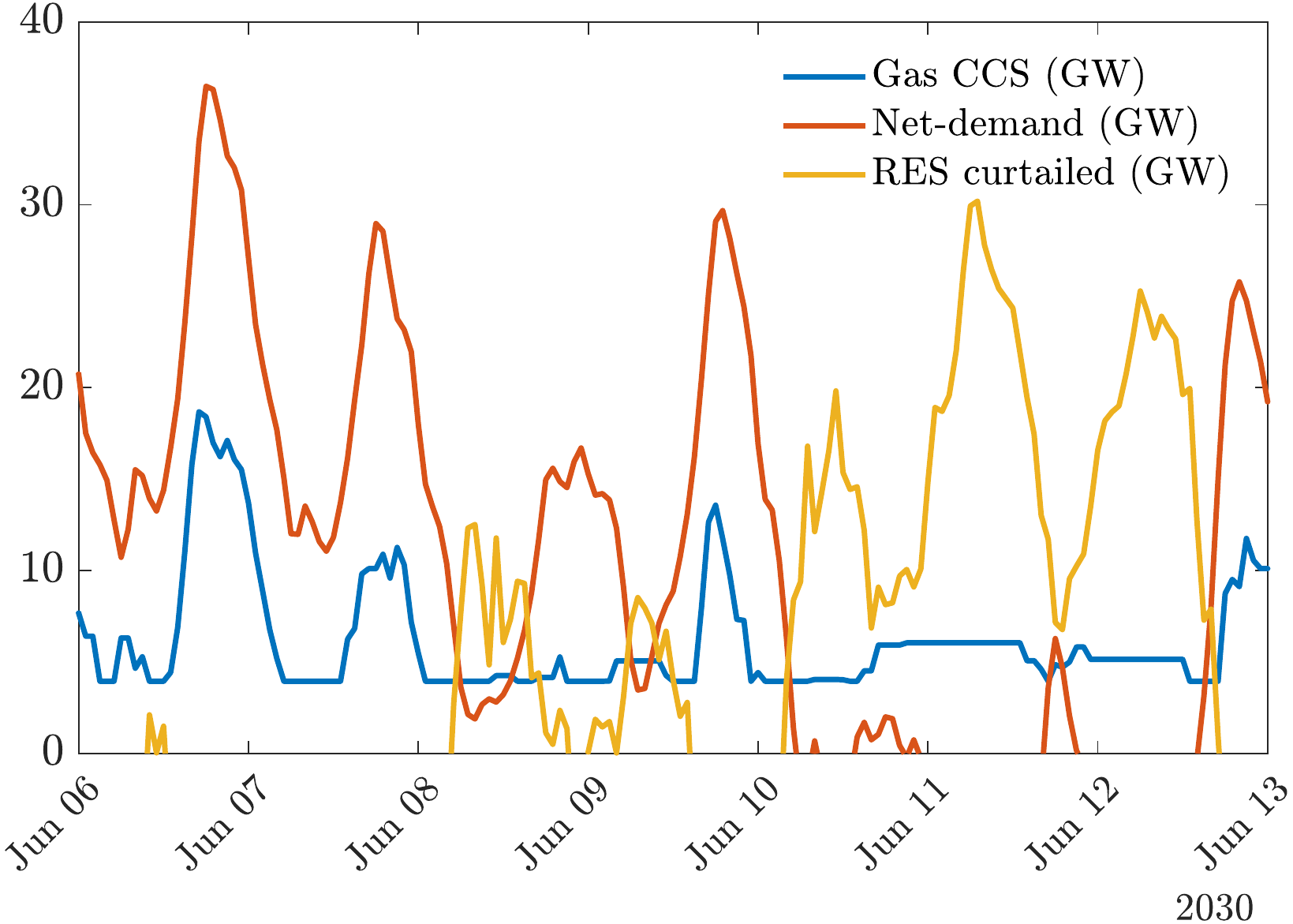}
    \caption{One week example of the projected system operation for 2030.}
	\label{fig:OneWeek_Operation}
\end{figure}

It is insightful to consider the distribution of system inertia throughout the year, as presented in Figure~\ref{fig:inertia_distribution}. In 2016, $220{\textrm{GVA}\cdot\textrm{s}}$ was the most common value \cite{NGrocofDefinition}, therefore inertia-related actions taken by the system operator were rare. On the other hand, the simulation results for 2030 show a distribution of inertia skewed towards significantly lower values. This distribution contains three peaks corresponding to the most common values of system inertia throughout the year:
\begin{itemize}
    \item Peak 1, at $74{\textrm{GVA}\cdot\textrm{s}}$, corresponds to hours for which only nuclear and gas plants are online providing a baseline of inertia, in order to respect the nadir limit given by eq.~(\ref{eq:nadir}). Although biomass plants have lower marginal costs than gas plants in Table~\ref{tab:GenerationMix}, the minimum stable generation of gas plants make them preferable in these hours as a source of inertia: this lower minimum stable generation allows to accommodate more energy from RES, which are assumed to have zero marginal costs, therefore reducing overall system costs.
    \item Peak 2, at $92{\textrm{GVA}\cdot\textrm{s}}$, corresponds to periods of high RES output when it is effective to commit a combination of biomass plants to provide inertia (since the inertia provided by biomass plants is cheaper than that from gas plants, given the lower marginal costs of biomass units), and some gas plants to provide the necessary PFR. 
    \item Peak 3, at $165{\textrm{GVA}\cdot\textrm{s}}$, corresponds to periods of high net-demand, when dispatchable, synchronous plants must be turned on for energy purposes, and therefore inertia effectively becomes a by-product of energy.
\end{itemize}

\begin{figure} [!t]
\hspace*{-1.5mm}
    \includegraphics[width=3.3in]{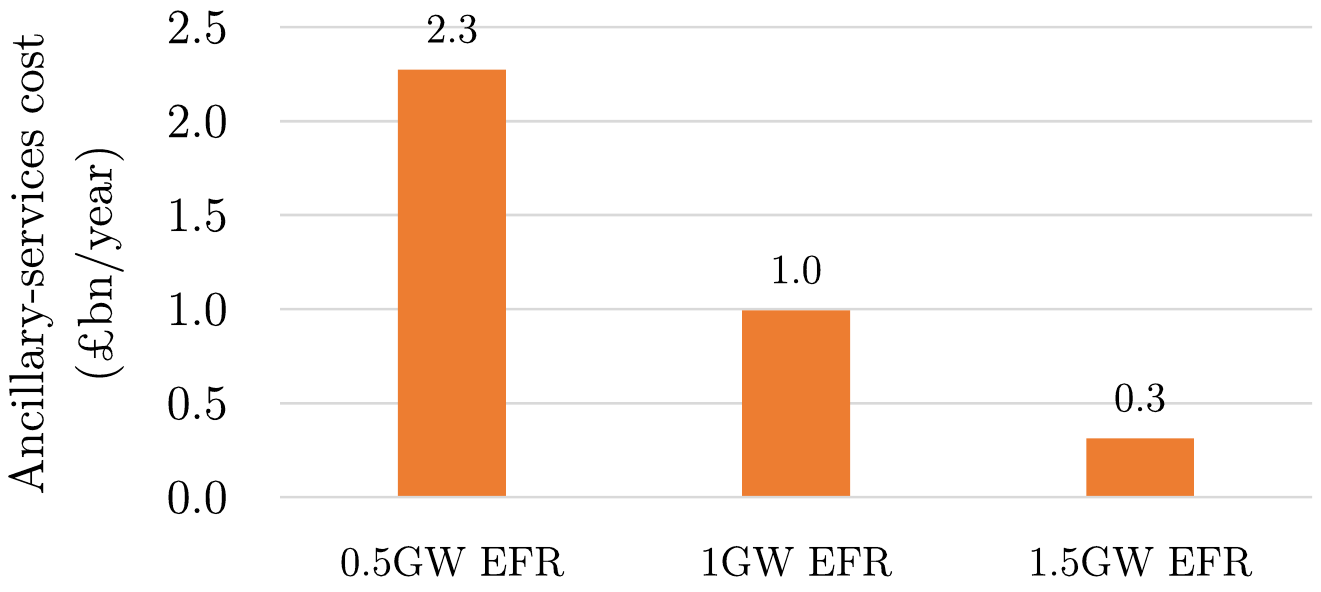}
    \caption{Projected cost of ancillary services for 2030, considering three different cases of volumes of EFR procured.}
    \label{fig:EFRsensitivity}
\end{figure}

To further understand the annual distribution of inertia for 2030 presented in Figure~\ref{fig:inertia_distribution}, an example of a week of operation of the system is included in Figure~\ref{fig:OneWeek_Operation}. The first two days of this week show periods of high net-demand, when the gas plants increase their output to cover the system demand. On the other hand, the last three days show periods of zero net-demand, when RES curtailment is present for a sustained period. Even though in these days all the demand could be covered by non-synchronous RES generation, a minimum amount of gas plants is kept online to provide inertia: in the current GB system, this would be achieved by bringing online out-of-merit thermal units through the Balancing Mechanism, in order to increase inertia and therefore guarantee system stability.

On a final note, it is important to clarify that RES do not provide ancillary services in the studies presented in this paper. That is currently the case in most countries in the world, as significant market barriers exist for these generators. Notably, ancillary services markets in GB are currently cleared well ahead of real time, which makes it impossible for RES to accurately forecast their potential contribution to these services. In addition, the current RES fleets, which are based on grid-following inverters, are widely agreed to be less effective than synchronous machines to contain frequency drops. Nevertheless, system operators are working to remove market barriers, and the adoption of grid-forming inverters could remove the technical limitations. Once RES are fully able to provide ancillary services, the stability challenge in decarbonised grids could be reduced, but the trend of increasing overall costs for these services projected in this paper is still likely to occur.

\begin{figure} [!t]
\hspace*{-1.5mm}
    \includegraphics[width=3.3in]{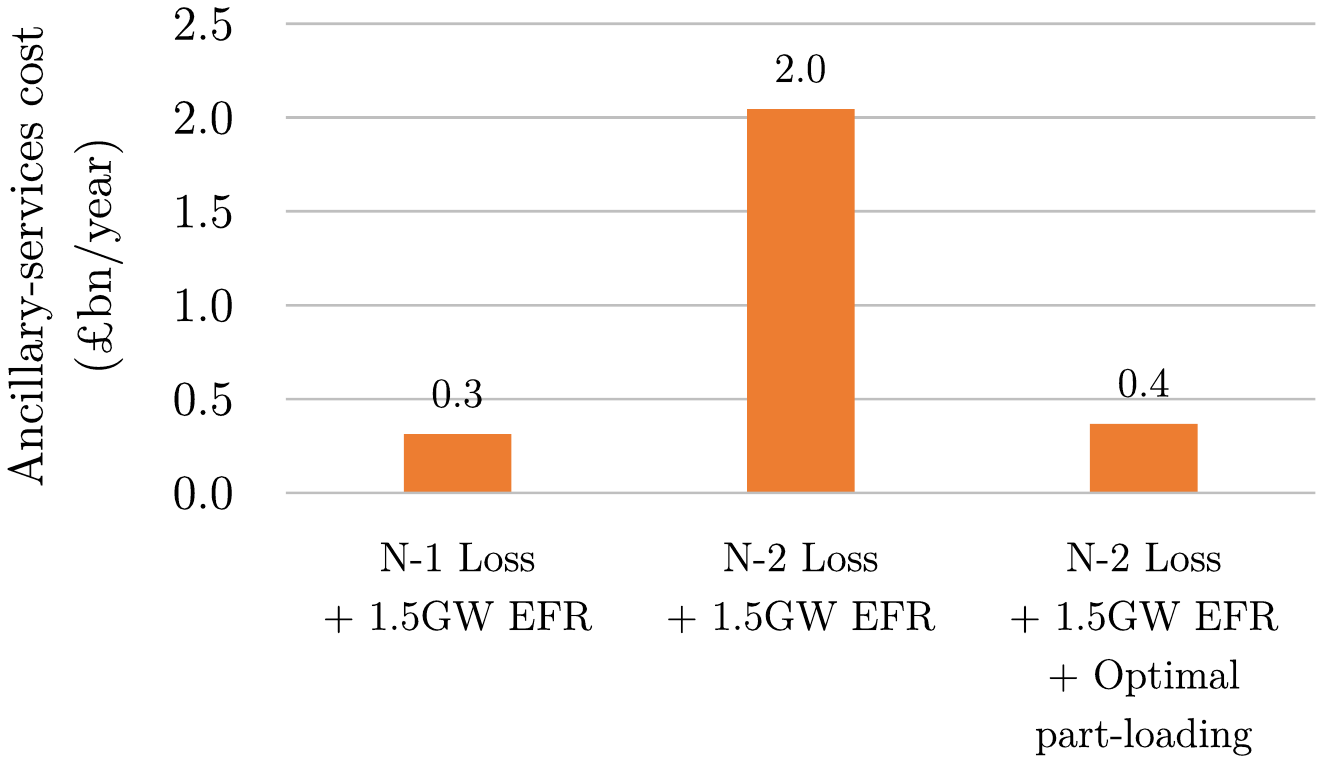}
    \caption{Projected cost of ancillary services for 2030, considering three different cases for the reliability standard enforced.}
	\label{fig:Deloading}
\end{figure}

\subsection{Fast frequency response to compensate for low inertia}

One of the keys to tackle the low-inertia challenge is the procurement of fast frequency-response services (such as EFR in GB), as demonstrated by works like \cite{LuisEFR}. Here we run a sensitivity analysis for the volume of EFR procured by 2030, for which the results are presented in Figure~\ref{fig:EFRsensitivity}. These results demonstrate that the cost of ancillary services could more than double if the volume of EFR procured is halved from the currently planned 1GW, reaching a share of total system costs of 35\% in this case (since projected energy costs for 2030 are of \pounds6.5bn/year, as presented in Figure~\ref{fig:ResultsSUC}). On the other hand, procuring an additional 0.5GW of EFR (making the total volume of 1.5GW) would reduce ancillary-services costs by roughly \pounds700m/year, highlighting the system value of this service. Note that during the 2020 lockdown, 200MW of EFR were being procured \cite{NationalGridEFR}. 

It is important to clarify that, since EFR is provided by BESS which do not incur fuel costs, the only system operating cost associated to EFR is a lost opportunity to use the BESS fully for energy arbitrage. In practice, BESS would be remunerated for providing EFR with a certain price set in an auction for this service. This revenue, along with other revenue streams such as the price differentials from doing arbitrage in the energy market, would allow BESS owners to recover the investment cost on these assets. Note however that the scheduling model used in this paper only considers the operational time-scale of the power system, assuming a fixed capacity mix, and therefore investment costs are out of the scope of this analysis.

\subsection{Impact of largest loss and reliability standards} \label{Sec:LargestLoss}

In previous sections, the largest loss infeed driven by the large nuclear plant Hinkley Point C has been considered to be fixed, i.e.~$P_\textrm{Loss}=1.8\textrm{GW}$. Here, we consider two alternative cases: 
\begin{itemize}
    \item The system is secured against an `\textit{N}-2' loss, which considers the simultaneous loss of Hinkley Point C power station and half of the 2GW rated capacity of the double-circuit HVDC interconnector IFA. The value of $P_\textrm{Loss}$ is then set to 2.8GW.
    \item A variation of the above case, where the size of the largest possible loss is optimised in the scheduling model, by optimally part-loading the large nuclear station if beneficial for the system as a whole. To compute this optimal part-loading strategy, $P_\textrm{Loss}$ is considered a decision variable in constraints~(\ref{eq:RoCoF}) and~(\ref{eq:nadir}), while respecting the minimum stable generation for Hinkley Point C in Table~\ref{tab:GenerationMix}. 
\end{itemize}
The results for these cases are presented in Figure~\ref{fig:Deloading}. A volume of 1.5GW of EFR is considered to be procured in all three cases, since it was verified that 1GW of EFR would not be enough to secure the `\textit{N}-2' case, given the generation mix considered in Table~\ref{tab:GenerationMix}.

These studies are motivated by the need to understand the cost of a tighter reliability standard, since the most recent large power outage in GB (on the $9^\textrm{th}$ of August 2019) was caused by an `\textit{N}-2' outage \cite{OfgemOutage2019}. The results in Figure~\ref{fig:Deloading} demonstrate that securing the system against any two simultaneous losses increases the cost of ancillary services by more than 6 times, even in the presence of 1.5GW of EFR. On the other hand, optimally sizing the largest possible loss is shown to achieve very significant savings. However, in this third case Hinkley Point C station is chosen by the optimal scheduling to operate on average at 1.45GW throughout the year; therefore, these savings from optimal part-loading can only be considered as an ideal scenario, since energy and ancillary services are fully co-optimised in the scheduling simulation in this case. In practice, nuclear part-loading would be done ex-post (e.g.~through the Balancing Mechanism in GB, which can adjust the day-ahead energy market dispatch), since Hinkley Point C will benefit from a CfD and therefore is incentivised to produce as much energy as possible. This matter is further discussed in the next section.

\subsection{Future market arrangements for renewables and nuclear plants}

In addition to the increased need for ancillary services in a low-inertia system, the special market arrangements currently in place for most renewables in GB complicated the challenge during the 2020 lockdown. In coming years, renewables might still benefit from instruments such as Contracts for Difference (CfDs), as the business case for merchant renewables (i.e.~renewables that simply obtain revenue from the wholesale market) is still under debate, due to issues such as price cannibalisation (i.e.~renewables would not be able to capture high prices, given the correlation between wind or solar output at different locations, respectively). Under a CfD, renewable generators would still be incentivised to produce as much energy as possible, regardless of the level of demand. Nevertheless, the latest round of CfD auctions held in GB forbids any recipient of a CfD to bid negatively \cite{CfD2019}, which could help alleviate the renewable output during periods of very low demand (when energy prices could potentially be negative).

Furthermore, the large nuclear plant Hinkley Point~C, which will drive the largest loss in GB in coming years, will operate under a CfD with a strike price of \pounds92.5/MWh (in 2012 prices, to which inflation must be added) \cite{HinkleyPointC}, placing it at around 45\% above average wholesale day-ahead electricity prices in the last decade \cite{OfgemPrices2020}.
This implies that it will be incentivised to run at output roughly at all times, which would highly increase the need for ancillary services due to the large size of this plant driving the `\textit{N}-1' reliability requirement.  
Some studies have proved that part-loading large nuclear units when net-demand is low can significantly decrease system operating costs and reduce overall emissions \cite{LuisPESGM2018}, therefore mechanisms for making this strategy compatible with a CfD should be designed.

\subsection{Applicability of the results to other countries}

The results presented in this paper have been obtained by using the projected future British electricity system as a test case. As an island, Great Britain is already suffering the low-inertia problem induced by increasing renewable penetration, which was particularly acute during the COVID-19 lockdown. Other island nations such as Ireland and Australia are experiencing the same issue, as well as Texas, which has no synchronous interconnection to other grids in the continental United States. 

However, the results presented here could also be of relevance to system operators and stakeholders in the energy sector in other countries. While continental grids like the US and Europe currently benefit from sharing inertial support through AC transmission corridors, once renewable penetration is sufficiently high in all the interconnected countries, the low-inertia issue will also appear in continental grids. Therefore, the lessons drawn from the simulations conducted in this paper will be of interest for these systems.

Note that the present paper makes no assumption on the particular market arrangements for procuring ancillary services in GB: the cost of ancillary services presented here is simply the opportunity cost computed with the frequency-secured Unit Commitment, as explained in Sections~\ref{Sec:SchedulingModel} and \ref{Sec:Results}. The only British particularity in terms of ancillary services that is considered in this paper is the existence of the `fast frequency response' service (i.e.~EFR), which also exists with different terminology in Australia, Ireland and Texas. While this product does not yet exist in most countries, it is not unreasonable to expect that a similar product will eventually be created by other system operators to counteract the low-inertia problem. The advantage of defining a `fast frequency response' product is to recognise the faster dynamics of certain grid-connected assets such as battery storage, and the results presented here demonstrate that such service can bring significant cost savings.

For quantifying the value of and need for the different ancillary services in a given country, it might nonetheless be relevant to consider its particular market arrangements. A working group within the `Conseil international des grands réseaux électriques' (CIGRE), gathering experts from system operators and academia across the world, is in the process of collecting operational practices for ancillary-services procurement in different countries. The interested reader can refer to the technical brochure which summarises the findings of the working group, expected to be published in Q1 2021 under the title `Impact of high penetration of inverter-based generation on system inertia of networks'.

\section{Conclusion and future work} \label{Sec:Conclusion}
The lockdown measures following the COVID-19 outbreak in GB has provided some valuable insights on the practical challenges of operating a system with low net-demand. While these challenges had already been predicted, they were not expected to be present until some years in the future, as the penetration of renewables keeps increasing to eventually reach the legal requirement of net-zero emissions by 2050 imposed by the British parliament. The exceptional circumstances of the pandemic, and the subsequent depression in electric demand, have shown that these challenges are very much real.

This paper has demonstrated and discussed why these challenges will become increasingly relevant as the share of renewables in the system grows. While the difficulty in managing the GB system will likely be higher than in continental grids, given that it is an island where the low-inertia problem is already present, other countries will certainly experience similar issues in the future as they progressively move towards a carbon-free energy system.

Regarding future work that could lead to improved practices in procurement of ancillary services, it is notably relevant to understand how to make compatible the energy-based, financial-support mechanisms for certain generators (e.g.~Contracts for Difference) with the provision of inertia, frequency response and an optimised largest infeed loss. This issue is particularly key in GB, where the future driver of the largest possible loss will be a nuclear plant benefiting from a CfD, which could exacerbate the stability challenge during periods of low net-demand.

\section*{Acknowledgements}
The authors are grateful for the valuable support and funding received by the UK EPSRC project `Integrated Development of Low-Carbon Energy Systems' (IDLES, Grant EP/R045518/1) and EU H2020 TradeRES project (ID: 864276).

The first author is also grateful to Dr.~José Enrique ``Villica'' Villa Escusol for the conversation that sparked the writing of this paper.

\vspace*{-3mm}
\section*{}

\bibliography{Luis_PhD.bib}

\appendix
\section{Unit Commitment formulation} \label{app:UC}

The optimisation problem described in Section~\ref{Sec:SchedulingModel} is subject to the constraints included in this appendix. Decision variables are italicised to distinguish them from constants. For the sake of simplicity, the formulation presented here corresponds to the deterministic UC, while the interested reader can refer to \cite{AlexEfficient} for a detailed description of the generalisation of these constraints to be applied to the stochastic UC. 

The only binary decision variables in this formulation are $y_g^\textrm{st}$ and $y_g^\textrm{sd}$, while all other are continuous. 
The results presented in Section~\ref{Sec:Results} through Section~\ref{Sec:LargestLoss} have been obtained with these binary variables being relaxed, in order to decrease computational time.

The load balance constraint, assuming an hourly time-step in the scheduling optimisation, is determined by:
\begin{equation} 
\sum_{\forall g} P_{g}+\sum_{\forall s} \left( P_s^\textrm{d} - P_s^\textrm{c} \right) + \textrm{P}_\textrm{RES} - P_\textrm{RES}^\textrm{curt} = \textrm{D} - \textrm{D}_\textrm{shed}  
\end{equation}
Where RES curtailment is limited by these inequalities:
\begin{equation}
    0 \leq P^\textrm{curt}_\textrm{RES} \leq \textrm{P}_\textrm{RES}
\end{equation}

\subsection{Constraints for thermal units}
The thermal units described in Table~\ref{tab:GenerationMix} are subject to the following constraints. The power limits:
\begin{equation}
    y_g\cdot\textrm{P}^\textrm{msg}_g \leq P_g \leq y_g\cdot\textrm{P}^\textrm{max}_g
\end{equation}
The ramp limits:
\begin{equation}
    -\textrm{P}_g^\textrm{rr} \cdot y_g(t-1) \leq P_g(t)-P_g(t-1)
    \leq \textrm{P}_g^\textrm{rr} \cdot y_g(t)
\end{equation}
The commitment state of a unit is `on' if it was generating in the previous time-step or has started generating in the current time-step, unless it has been shut down in the current time-step:
\begin{equation}
    y_g(t) = y_g(t-1) + y_g^\textrm{sg}(t) - y_g^\textrm{sd}(t)
\end{equation}
The thermal unit starts generating if it was started up `$\textrm{T}_g^\textrm{st}$' hours before:
\begin{equation}
    y_g^\textrm{sg}(t) = y_g^\textrm{st}(t-\textrm{T}_g^\textrm{st}) 
\end{equation}
The generator is allowed to be started up if it was `off' in the previous time-step and has been `off' for at least `$\textrm{T}_g^\textrm{mdt}$' hours:
\begin{equation}
    y_g^\textrm{st}(t) \leq
    \left[1-y_g(t-1)\right] - \hspace{-1mm} \sum_{i=t-\textrm{T}_g^\textrm{mdt}}^t y_g^\textrm{sd}(i)
\end{equation}
The unit is allowed to be shut down if it was generating in the previous time-step, but also has been generating for at least `$\textrm{T}_g^\textrm{mut}$' hours:
\begin{equation}
    y_g^\textrm{sd}(t) \leq
    y_g(t-1) - \hspace{-1mm} \sum_{i=t-\textrm{T}_g^\textrm{mut}}^ty_g^\textrm{sg}(i)
\end{equation}

Regarding the contribution of a thermal unit to PFR, this is limited by its headroom and by the PFR capacity of the unit:
\begin{equation}
    \textrm{PFR}_g \leq y_g \cdot \textrm{P}_g^\textrm{max} - P_g 
\end{equation}
\vspace{-5mm}
\begin{equation} 
    \textrm{PFR}_g \leq \textrm{PFR}_g^\textrm{max}
\end{equation}

\subsection{Constraints for storage units}

PHES and BESS units are subject to the following constraints. The only binary variable is the charge/discharge switch, $y_s$, which prevents the storage unit from charging and discharging simultaneously. 
All other variables are continuous.

The energy stored is defined by:
\begin{equation}
    \textrm{E}_s^\textrm{min} \leq E_s \leq \textrm{E}_s^\textrm{max}
\end{equation}
\vspace{-6mm}
\begin{equation}
    E_s(t) = E_s(t-1) + \left( \eta_s^\textrm{c} \cdot P_s^\textrm{c}(t) - \frac{P_s^\textrm{d}(t)}{\eta_s^\textrm{d}} \right) 
\end{equation}
The charge and discharge power are constrained by:
\begin{equation}
    0 \leq P_s^\textrm{c} \leq (1-y_s) \cdot \textrm{P}_s^\textrm{c,max}
\end{equation}
\vspace{-5mm}
\begin{equation}
    0 \leq P_s^\textrm{d} \leq y_s \cdot \textrm{P}_s^\textrm{d,max}
\end{equation}

\begin{figure} [t!]
    \centering
    \includegraphics[width=3.3in]{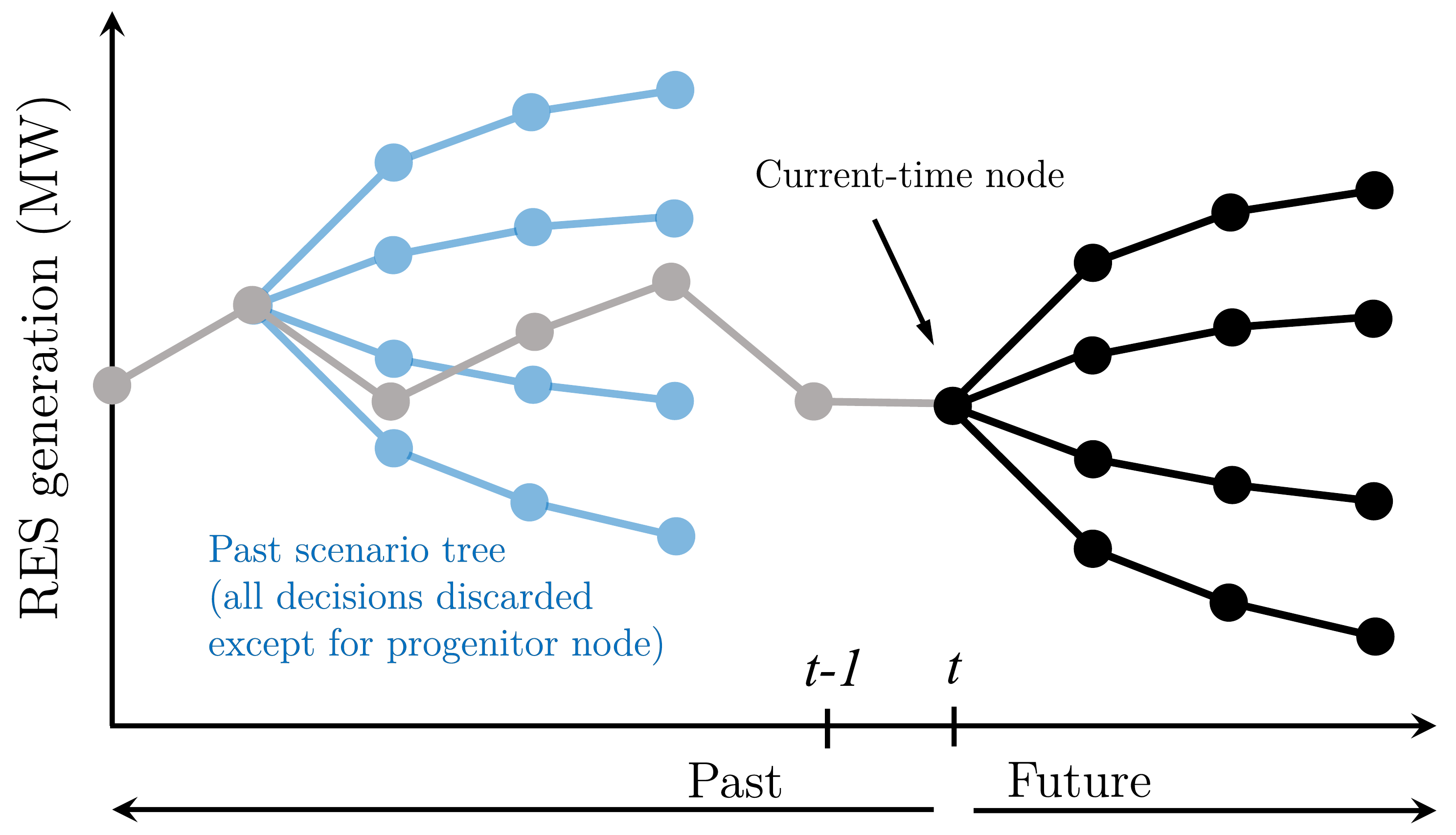}
    \caption{Schematic of the characterisation of uncertainty in RES generation within the stochastic UC.}
	\label{fig:ScenarioTree}
\end{figure}

The contribution of BESS to EFR is determined by:
\begin{equation} 
    \textrm{EFR}_s \leq \textrm{EFR}_s^\textrm{max}
\end{equation}
\vspace{-5mm}
\begin{equation}
    \textrm{EFR}_s \leq y_s \cdot \textrm{P}_s^\textrm{d,max} - P_s^\textrm{d} + P_s^\textrm{c} 
\end{equation}
The last constraint accounts for the fact that BESS can swiftly switch from charging to discharging, therefore effectively providing a higher volume of EFR if it operates in charging mode.

\subsection{Definitions of frequency-related magnitudes}

This frequency-secured UC guarantees frequency stability by including constraints (\ref{eq:RoCoF}) and (\ref{eq:nadir}). The expressions that appear in these two constraints are defined in this subsection.

The total system PFR is given by:
\begin{equation} 
    \textrm{PFR} = \sum_{\forall g} \textrm{PFR}_g 
\end{equation}
The total system EFR is defined by:
\begin{equation} 
    \textrm{EFR} = \sum_{\forall s} \textrm{EFR}_s 
\end{equation}

System inertia is the aggregate of the inertia from each thermal plant online at a given time:
\begin{equation} 
    H = \sum_{\forall g} \textrm{H}_g \cdot \textrm{P}_g^\text{max} \cdot y_g - \textrm{P}_\textrm{Loss}^\textrm{max} \cdot \textrm{H}_\textrm{Loss} 
\end{equation}
which accounts for the loss of inertia from the outaged generator.

The `fixed' largest loss is defined as:
\begin{equation} 
    P_\textrm{Loss} = \textrm{P}_\textrm{Loss}^\textrm{max}
\end{equation}
While the `optimised' largest loss, achieved by optimally part-loading large units in the scheduling, is determined by:
\begin{equation} 
    P_\textrm{Loss} = P_g \qquad \textrm{for} \; g=\textrm{Largest Unit}
\end{equation}
The above constraint assumes that the largest unit is a must-run unit, i.e.~it is committed at all times. This is the case for the GB system described in Table~\ref{tab:GenerationMix}, where the largest unit is the nuclear plant Hinkley Point C.

\vspace*{7mm}
\subsection{Uncertainty from RES} \label{app:UncertaintyUC}

Uncertainty in RES generation is included in the stochastic Unit Commitment through a scenario tree. This scenario tree discretises the probability density function of forecast errors for the RES generation. Full details on how to compute this discretisation and the probabilities of reaching each of the nodes, $\pi(n)$, can be found in \cite{AlexEfficient}.

Figure~\ref{fig:ScenarioTree} displays a generic example of the use of scenario trees within the stochastic scheduling. The stochastic UC is run with a 24h lookahead: at each time-step, a scenario tree is built for the following 24h hours. A rolling-planning approach is also used, consisting in only keeping the optimisation solution for the current-time node in the scenario tree (the progenitor of all nodes) and discarding all other decisions. This approach has been shown to provide lower cost solutions for systems with a high renewable penetration.

\end{document}